\documentclass[reprint, amsmath, amssymb, aps, prd, superscriptaddress, nofootinbib, nobibnotes,longbibliography]{revtex4-1}
\usepackage{graphicx}
\usepackage{subfig}

\usepackage{bm}
\usepackage{multirow}
\usepackage{amsmath}
\usepackage[colorinlistoftodos]{todonotes}
\usepackage[colorlinks=true, allcolors=blue]{hyperref}
\graphicspath{{figures/}}
\usepackage{acro}
\DeclareAcronym{EMRI}{
	short = EMRI,
	long  = extreme mass ratio inspiral
}
\DeclareAcronym{PTA}{
	short = PTA,
	long  = pulsar timing array
}

\DeclareAcronym{LISA}{
	short = LISA,
	long  = Laser Interferometer Space Antenna
}

\DeclareAcronym{GR}{
	short = GR,
	long  = general relativity
	}
	
\DeclareAcronym{BH}{
	short = BH ,
	long  = black hole
}
\DeclareAcronym{PBH}{
	short = PBH ,
	long  = primordial black hole
}
\DeclareAcronym{SGWB}{
	short = SGWB ,
	long  = stochastic gravitational-wave background
}
\DeclareAcronym{BBH}{
	short = BBH ,
	long  = binary black hole
}
\DeclareAcronym{BNS}{
	short = BNS ,
	long  = binary neutron star
}
\DeclareAcronym{NSBH}{
	short = NSBH ,
	long  = neutron star black hole
}

\DeclareAcronym{GW}{
	short = GW ,
	long  = gravitational wave
}
\DeclareAcronym{CMB}{
	short = CMB ,
	long  = cosmic microwave background
} 
\DeclareAcronym{SFR}{
	short = SFR ,
	long  = star formation rate
}

\DeclareAcronym{CBC}{
	short = CBC,
	long  = compact binary coalescence
}
\DeclareAcronym{SNR}{
	short = SNR,
	long  = signal noise ratio
}
\DeclareAcronym{IMR}{
	short = IMR,
	long  = inspiral-merger-ringdown
}

\DeclareAcronym{PSD}{
	short = PSD,
	long  = power density spectrum
}

\DeclareAcronym{QNM}{
	short = QNM,
	long  = quasi-normal modes
}

\DeclareAcronym{GRB}{
	short = GRB,
	long  = gamma-ray burst
}

\DeclareAcronym{FLRW}{
	short = FLRW,
	long  = Friedmann-Lema\^{i}tre-Robertson-Walker
}

\usepackage[capitalise]{cleveref}
\crefname{figure}{Fig.}{Figs.}
\Crefname{figure}{Fig.}{Figs.}

\def\be{\begin{equation}}
\def\ee{\end{equation}}
\def\({\left(}
\def\){\right)}
\def\[{\left[}
\def\]{\right]}

\begin{document}

\title{Searching for primordial black holes with stochastic gravitational-wave background in the space-based detector frequency band}
\author{Yi-Fan Wang}
\email{yifan.wang@aei.mpg.de}
\affiliation{Max-Planck-Institut f{\"u}r Gravitationsphysik (Albert-Einstein-Institut), D-30167 Hannover, Germany}
\affiliation{Leibniz Universit{\"a}t Hannover, D-30167 Hannover, Germany}
\affiliation{Department of Physics, The Chinese University of Hong Kong, Shatin, New Territories, Hong Kong}
\author{Qing-Guo Huang}
\affiliation{CAS Key Laboratory of Theoretical Physics, Institute of Theoretical Physics,Chinese Academy of Sciences, Beijing 100190, China}
\affiliation{School of Physical Sciences, University of Chinese Academy of Sciences, No. 19A Yuquan Road, Beijing 100049, China}
\affiliation{Center for Gravitation and Cosmology, College of Physical Science and Technology,Yangzhou University, 88 South University Avenue, 225009, Yangzhou, China}
\affiliation{Synergetic Innovation Center for Quantum Effects and Applications,Hunan Normal University, 36 Lushan Lu, 410081, Changsha, China}
\author{Tjonnie G.F. Li}
\affiliation{Department of Physics, The Chinese University of Hong Kong, Shatin, New Territories, Hong Kong}
\author{Shihong Liao}
\affiliation{Key Laboratory for Computational Astrophysics, National Astronomical Observatories, Chinese Academy of Sciences, Beijing 100012, China}

\begin{abstract}
Assuming that primordial \aclp{BH} compose a fraction of dark matter, some of them may accumulate  at the center of galaxy and perform a prograde or retrograde orbit against the gravity pointing towards the center exerted by the central massive \acl{BH}.
If the mass of primordial \aclp{BH} is of the order of stellar mass or smaller, such \aclp{EMRI} can emit \aclp{GW} and form a background due to incoherent superposition of all the contributions of the Universe. 
We investigate the \acl{SGWB} energy density spectra from the directional source, the primordial \aclp{BH} surrounding Sagittarius A$^\ast$ of the Milky Way, and the isotropic extragalactic total contribution, respectively.
As will be shown, the resultant \acl{SGWB} energy density shows different spectrum features such as the peak positions in the frequency domain for the above two kinds of sources.
Detection of \acl{SGWB} with such a feature may provide evidence for the existence of primordial \aclp{BH}. 
Conversely, a null searching result can put constraints on the abundance of primordial \aclp{BH} in dark matter.
\end{abstract}

\maketitle
\acresetall

\section{Introduction} 

The recent direct detections of \acp{GW} by the LIGO and Virgo collaborations open a unique window to observe \acp{BH}
\cite{Abbott:2016blz,Abbott:2016nmj,O1,Abbott:2017vtc,Abbott:2017oio,TheLIGOScientific:2017qsa,Abbott:2017gyy,GWTC1}.
The event rate of binary \ac{BH} merger at local Universe is estimated to be $53.2^{+58.5}_{-28.8}$ Gpc$^{-3}$ yr$^{-1}$ from the detections \cite{GWTC1-rate}. 
Among the \ac{GW} events, the relatively large mass of the first detection ($\sim 30 M_\odot$), GW150914, has stimulated discussions that the binary \acp{BH} of GW150914 could be of primordial origin, instead of products of stellar evolution \cite{Bird:2016dcv,Clesse:2016vqa,Sasaki:2016jop,Chen:2018czv}.
Ref.~\cite{Sasaki:2016jop} shows that the binary stellar-mass primordial \acp{BH} coalescence scenario can give the correct order of magnitude of event rate, if the abundance of primordial \acp{BH} in dark matter is $\sim10^{-3}$.

Primordial \acp{BH} are a long hypothesized candidate for dark matter \cite{Hawking:1971ei,Carr:1974nx,PBH1993-Silk,PBH1996}. 
Assuming all the primordial \acp{BH} have the same mass, a variety of observations from astronomy and cosmology have given constraints on the primordial \ac{BH} abundance in dark matter, for example, gravitational lensing of stars and quasars, dynamics of dwarf galaxies, large scale structure formation and accretion effects on the \ac{CMB} (see Refs.~\cite{Carr:2016drx,Sasaki:2018dmp,Belotsky:2018wph,Ketov:2019mfc} and references therein). 
The possibility that all the dark matter are primordial \acp{BH} with the same mass has been ruled out given all the constraints aforementioned (but see, e.g., Ref.~\cite{Montero-Camacho:2019jte}).
Nevertheless, it is still interesting to consider the scenario where primordial \acp{BH} compose a part of dark matter and propose new methods to seek for evidence of primordial \acp{BH} or constrain their abundance in dark matter, especially leveraging the newly opened \ac{GW} window \cite{LIGO-sub-solar-O1, LIGO-sub-solar-O2}.

In this work, we investigate the scenario in which primordial \acp{BH} constitute a fraction of dark matter in the galactic center.
Astrophysical observations (see, e.g., Refs.~\cite{SMBH-Review-2,SMBH-Review-1} for a review) indicate that massive \acp{BH} with mass $10^5 M_\odot- 10^{9} M_\odot$ are ubiquitous and reside at the center of almost every massive galaxy. 
If some fraction of dark matter is composited by  primordial \acp{BH}, they should perform a prograde or retrograde orbit against the gravity pointing towards the galactic center exerted by the central massive \ac{BH}, and such a system becomes the so-called \ac{EMRI} system whose mass ratio is usually larger than $10^5$ \cite{Babak:2017tow}.
\acp{EMRI} are one of the important scientific targets of space-based \ac{GW} detector, such as \ac{LISA} which is anticipated to be launched in the 2030s \footnote{https://www.elisascience.org/}.
Once detected, the \ac{GW} signals from \acp{EMRI} can provide valuable information such as event rate estimation \cite{Babak:2017tow,AmaroSeoane:2010bq} and tests of general relativity \cite{Barack:2006pq,Gair:2012nm}.

The focus of our work is the \ac{SGWB} energy density spectrum from the \ac{EMRI} system consisting of a massive \ac{BH} at the galactic center and a subsolar mass primordial \ac{BH}.   
\ac{SGWB} is an incoherent superposition of numerous \acp{GW}, including those too weak to be detected individually \cite{Regimbau:2011rp,SGWBlivingreview,TheLIGOScientific:2016wyq,TheLIGOScientific:2016dpb,Abbott:2017xzg} or having an intrinsic stochastic nature, such as the primordial \ac{GW} which is generated by quantum fluctuations in the early Universe. 
The \ac{SGWB} from binary stellar-mass primordial \ac{BH} coalescence is calculated by Refs.~\cite{Mandic:2016lcn,Wang:2016ana,Chen:2018rzo,Clesse:2016ajp} and, in particular, Ref.~\cite{Wang:2016ana} shows that the null result of \ac{SGWB} in the LIGO frequency band ($\sim[10,1000]$ Hz) has given the most stringent constraints on the abundance of primordial \ac{BH} as dark matter in the mass range $[1,100] ~M_\odot$.
In the \ac{LISA} frequency band, 
Refs.~\cite{Kuhnel:2017bvu, SgrA} have considered the stochastic background from subsolar mass primordial \acp{BH} inspiraling to Sagittarius A$^\ast$, i.e., the central massive \ac{BH} of the Milky Way. 
Our work will expand the study of Refs.~\cite{Kuhnel:2017bvu, SgrA} in the following two aspects. First, we  calculate the \ac{SGWB} energy density spectrum in the frequency domain, i.e., $\Omega_\text{GW}(\nu)$ from the primordial \acp{BH} surrounding Sagittarius A$^\ast$. Second, we investigate the complete \ac{SGWB} contributions from extragalactic sources by modeling the event rate of primordial \ac{BH} \acp{EMRI} throughout the cosmic redshift.

The rest of the paper is arranged as follows.
In \cref{sec:sgrA} we model the primordial \acp{BH} density profile around a massive \ac{BH}, and apply this relation to the Sagittarius A$^\ast$ in the Milky Way to derive the \ac{SGWB} energy density spectrum.
We proceed to model the number density of massive \acp{BH} at different redshift epochs and calculate the \ac{SGWB} spectrum from extragalactic sources in \cref{sec:extragalactic}.
We forecast the ability of \ac{LISA} for detecting the \ac{SGWB} signal or, if there is a null result, constraining the abundance of primordial \acp{BH} in dark matter in \cref{sec:PBH-EMRI-constraint}.
The results show that \ac{LISA} can probe the existence of primordial \acp{BH} with mass range $[10^{-8} , 1 ]~M_\odot$ and constrain the abundance of primordial \ac{BH} with $1 M_\odot$ to be $10^{-9}$ in the optimal case where the dark-matter spike scenario with a steeper initial power index $\gamma=2$ is valid.
The main uncertainty is subject to the value of the dark-matter initial power index $\gamma$.
We summarize the conclusions in \cref{sec:PBH-EMRI-conclusion}.
Throughout this work we assume the mass distribution of primordial \acp{BH} is a delta function due to the uncertainty of the primordial \ac{BH} population.
Therefore the results should be seen as being from the primordial \acp{BH} with a representative mass.
Actually, as will be shown in the following, the mass of primordial \acp{BH} only serves as a scaling factor for the amplitude of the resulting \ac{SGWB} spectra and the shape of spectra only depends on the mass of the central massive \acp{BH}.

\section{The Stochastic Gravitational-Wave Background from Sagittarius A$^\ast$}\label{sec:sgrA}
\subsection{Primordial Black Holes Density Profile} 
To model the event rate of primordial \ac{BH} \acp{EMRI}, we first infer the primordial \ac{BH} mass density at the galactic center. 
Since we expect that primordial \acp{BH} compose a part of dark matter, it is natural to use the dark-matter density profile to characterize the primordial \ac{BH} mass density around the central massive \ac{BH}.

For an initial dark-matter density profile with the following power law form
\be\label{Eq:powerindex}
\rho(r) = \rho_0\left(\frac{r_0}{r}\right)^{\gamma},
\ee
where $\rho_0$ and $r_0$ are halo parameters and to be determined, $\gamma$ is the power index, $r$ is the radius of dark matter, Ref.~\cite{PhysRevLett.83.1719} suggests that the adiabatic growth of  the central massive \ac{BH} can enhance the surrounding dark-matter density at galactic center and form a spike distribution, i.e., the halo will end up with the following density \cite{PhysRevLett.83.1719,Nishikawa:2017chy}
\be \label{eq:spike}
\rho_\text{sp}(r) = \rho_R \left( 1-\frac{4R_s}{r}\right)^3\left(\frac{R_\text{sp}}{r}\right)^{\gamma_\text{sp}},
\ee 
where the power index is enhanced from the initial value by $\gamma_\text{sp} = (9-2\gamma)/(4-\gamma)$, the halo parameter $\rho_R = \rho_0(R_\text{sp}/r_0)^{-\gamma}$ with $R_s$ being the Schwarzschild radius of the central massive \ac{BH}, $R_\text{sp}(\gamma,M_\text{MBH}) = \alpha_\gamma r_0(M_\text{MBH}/(\rho_0 r_0^3))^{1/(3-\gamma)}$ is the radius to which the dark-matter spike extends, $\alpha_\gamma$ is derived numerically for different $\gamma$ in Ref.~\cite{PhysRevLett.83.1719}. 
For an initial Navarro-Frenk-White (NFW) profile \cite{NFW} with $\gamma=1$, the final spike has a index $\gamma_{\textrm{sp}}=7/3$, thus significantly boosting the inner profile around the central massive \ac{BH}.

To connect $\rho_0$ and $r_0$ with the massive \ac{BH}'s property, we employ the relation among the dark-matter halo virial mass $M_\text{vir}$, the concentration parameter $c_\text{con}\equiv r_\text{vir}/r_0$ where $r_\text{vir}$ is the halo virial radius, and the mass of massive \ac{BH} $M_\text{MBH}$. 
The relation of $c_\text{con}-M_\text{vir}$ for NFW profile is given by Ref. \cite{Dutton:2014xda}
\begin{equation}\label{eq:c}
\log {c_\text{con}} = a + b \log{ \frac{M_\text{vir}}{10^{12}h^{-1}M_\odot}    },
\end{equation}
where $a = 0.520+0.385\exp(-0.617z^{1.21})$ and $b = -0.101 + 0.026z$ are numerical factors at redshift $z$.
The parametrized formula \cref{eq:c} is obtained by numerically fitting to a suite of N-body simulations for NFW profile with \textit{Planck} 2013 cosmological parameters \cite{Planck2013} in the redshift range $[0,5]$.

The mass of the central massive \acp{BH} has a correlation with a few characteristic quantities of the host galaxy, such as the velocity dispersion $\sigma$ in the spheroidal region and the total mass of the host galaxy, indicating a coevolution history with the whole galaxy. 
By employing the observational $M_\textrm{MBH}-\sigma$ relation and using  the quasar luminosity function to link $\sigma$ with the halo mass, Ref.~\cite{Croton:2009zx} gives a parametrized relation between the massive \ac{BH}'s mass $M_\textrm{MBH}$ and the host dark-matter halo's virial mass $M_\text{vir}$,
\begin{align}
\log \left( \frac{M_\text{MBH}}{10^8 h^{-1} M_\odot}\right) &= (-2.66\pm 0.33) +(1.39 \pm  0.22) \nonumber \\
 &\times \log 	\left[  \beta^3  H(z) \left( \frac{M_\text{vir}}{10^{13} h^{-1} M_\odot }\right)\right].
 \label{Eq:MBH}
\end{align}
Here $H(z)=H_0 \sqrt{\Omega_m(1+z)^3+\Omega_\Lambda}$ is the Hubble parameter at redshift $z$ , $\Omega_m$ and $\Omega_\Lambda$ are the matter and dark energy fractional densities, respectively. $\beta$ is a ratio between the dark-matter halo's circular velocity and virial velocity, whose value is of order unity.
\cref{eq:c} together with \cref{Eq:MBH} can fix the corresponding coefficients of NFW density profile and NFW induced spike profile given the mass of massive \ac{BH}.
 
  \begin{figure}[htbp] 
   \centering
   \includegraphics[width=\columnwidth]{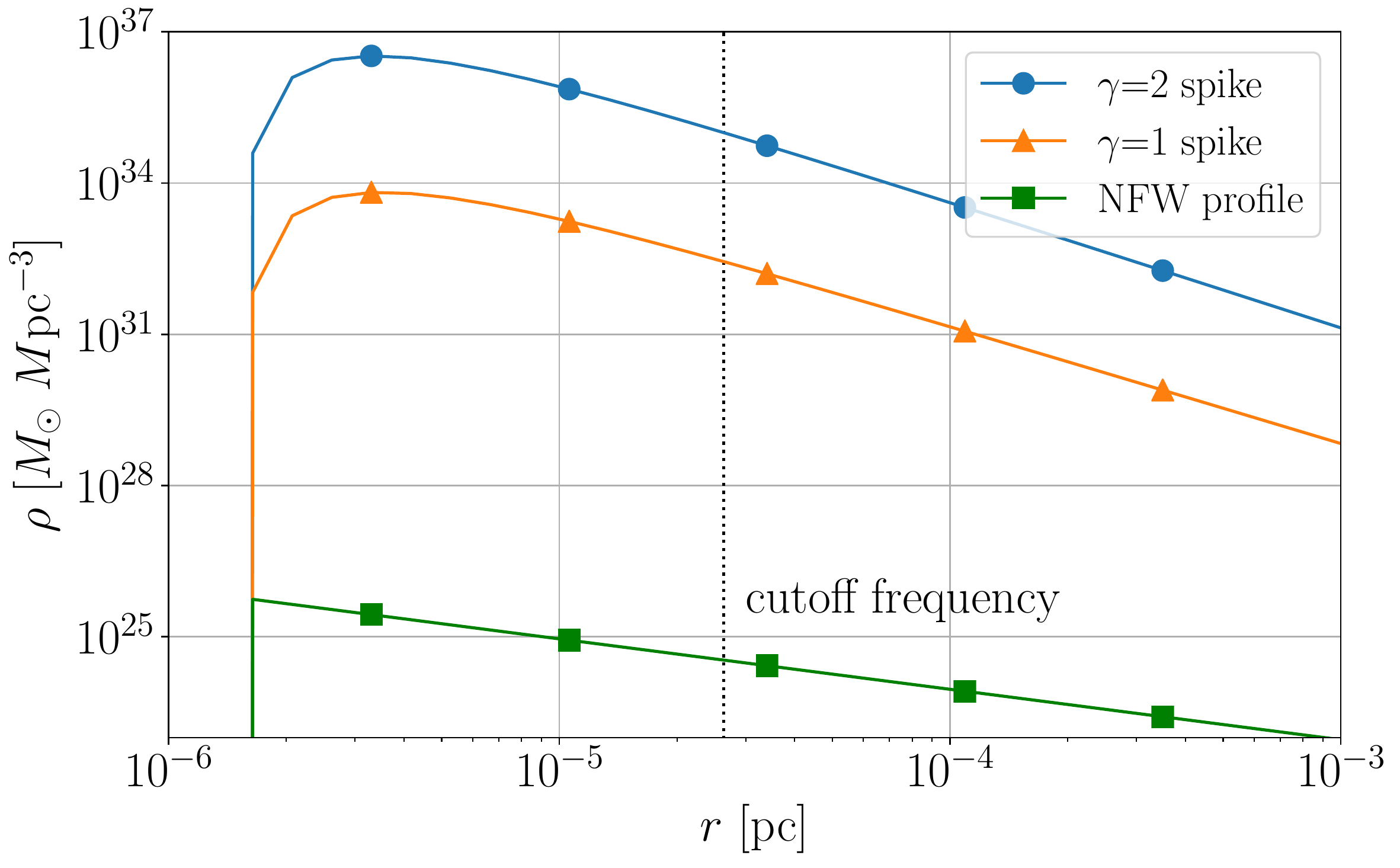} 
   \caption{Dark matter density profile around a massive \ac{BH} with mass $4\times 10^6 M_\odot$. The blue and orange solid curves show the spike profile for power index $\gamma = 2$ and $\gamma=1$, respectively. 
 As a comparison, the NFW dark-matter profile is also plotted.
The dark-matter density profile at galactic center is  boosted significantly for the $\gamma=1$ spike profile compared with the NFW profile. 
The $\gamma=2$ spike has even larger density than $\gamma=1$ by three orders of magnitude.
The vertical dashed line represents the orbital radius where the \ac{GW} frequency is $10^{-5} $Hz which is the lower cutoff frequency of the LISA sensitive band.
}
   \label{fig:profile}
\end{figure}

\cref{fig:profile} shows the dark-matter spike profile around a massive \ac{BH} with mass $4\times 10^6 M_\odot$.
Given the uncertainty of the initial profile power index, we choose two values for $\gamma$, $\gamma=1$ for NFW profile and $\gamma=2$ representing a steeper profile for an optimistic result.
We observe that the dark-matter density near the galactic center is  boosted significantly for the $\gamma=1$ spike profile compared with the NFW profile. 
The $\gamma=2$ spike has even larger density than $\gamma=1$ by three orders of magnitude.

In \cref{fig:profile}, the vertical dashed line represents the radius where the emitted \ac{GW} frequency is $10^{-5}$ Hz where we have assumed primordial \acp{BH} perform circular motion.
Therefore, the region of interest for the LISA frequency band is between $4 R_s$ as indicated by \cref{eq:spike}, and the dashed line.

 \subsection{Stochastic Gravitational-Wave Background Spectrum}

The fractional energy density spectrum of \ac{SGWB} is defined as
\begin{equation}
\Omega_\textrm{GW} = \frac{\nu}{\rho_c}\frac{d\rho_\textrm{GW}}{d\nu},
\end{equation}
where $d\rho_\textrm{GW}$ is the gravitational-wave energy density in the frequency band $[\nu,\nu+d\nu]$, $\rho_c=3H_0^2c^2/8\pi G$ is the critical energy to close the Universe, $G$ is the gravitational constant, $c$ is the speed of light.

The local energy density is related to the \ac{GW} energy flux by $\rho_\text{GW}(\nu) = F_\text{GW}(\nu)/c$.
$F_\text{GW}(\nu)$ from Sagittarius A$^\ast$ can be derived by taking the integral of the density of primordial \ac{BH} \ac{EMRI} with respect to the orbital radius $r$,
\begin{equation} \label{Eq:Fgw}
F_\text{GW}(\nu) = \int{f_\text{PBH}\rho_\text{DM}(r;M_\text{MBH})\over m_\text{PBH}} {dE/dt(r;M_{tot},\eta) \over {d_L^2}}r^2 dr, 
\end{equation}
where $d_L$ is the luminosity distance from sources to \ac{GW} detectors, $\rho_\text{DM}(r;M_\text{MBH})$ is the dark-matter density  at radius $r$ to the center given $M_\text{MBH}$, $m_\text{PBH}$ and $f_\text{PBH}$ are the primordial \ac{BH} mass and abundance, $\eta\equiv m_\text{PBH} M_\text{MBH}/(m_\text{PBH}+M_\text{MBH})^2$ is the symmetric mass ratio, $M_{tot}\equiv m_\text{PBH}+M_\text{MBH}$ is the total mass of the primordial \ac{BH} \ac{EMRI} system.
To the leading order, \ac{GW} power can be calculated by the quadrupole formula \cite{peter1,peter2},
\be
\frac{dE}{dt}(r;M_{tot},\eta)=\frac{32}{5}\frac{G^4}{c^5}\eta^2\left(\frac{M_{tot}}{r}\right)^5.
\ee
The radius $r$ can be equivalently replaced by the \ac{GW} frequency $\nu$ under the assumption of Keplerian motion,
\be\label{Eq:Kepler}
r(\nu,M_{tot})=\left[{(GM_{tot})^{1/2}\over \pi (1+z)\nu}\right]^{2\over 3}, 
\ee
where the factor of $(1+z)$ accounts for the cosmological expansion and the \ac{GW} frequency $\nu$ is twice of the orbital frequency.
From \cref{Eq:Kepler} a differential relation can  be derived ,
\be\label{Eq:differential}
{d\over d\nu}=-{2\pi\over 3}{1+z\over (GM_{tot})^{1/2}} r^{5/2}{d\over dr}.
\ee
Applying \cref{Eq:differential} to \cref{Eq:Fgw} and using the condition $M_\text{MBH}\gg m_\text{PBH}$, the differential \ac{GW} power from Sagittarius A$^\ast$ is
\be\label{Eq:dFoverdnu}
\frac{dF_\text{GW}}{d\nu}(\nu) = \frac{64\pi}{15} \frac{(1+z)G^{7/2}}{c^5}\frac{f_\text{PBH}\rho_\text{DM}m_\text{PBH}M_\text{MBH}^{5/2}}{r^{1/2}d_L^2},
\ee
Therefore the \ac{SGWB} energy density spectrum is
\begin{equation}\label{Eq:SgrA}
\Omega_\textrm{GW}^{Sgr A^{\ast}}(\nu) = \frac{\nu}{\rho_c}\frac{64\pi G^{7/2}}{15c^6} \frac{f_\textrm{PBH} m_\textrm{PBH} M_\text{MBH}^{5/2} \rho_\textrm{DM}}{ r^{1/2} d_L^2}.
\end{equation}

\begin{figure}[htbp] 
   \centering
   \includegraphics[width=\columnwidth]{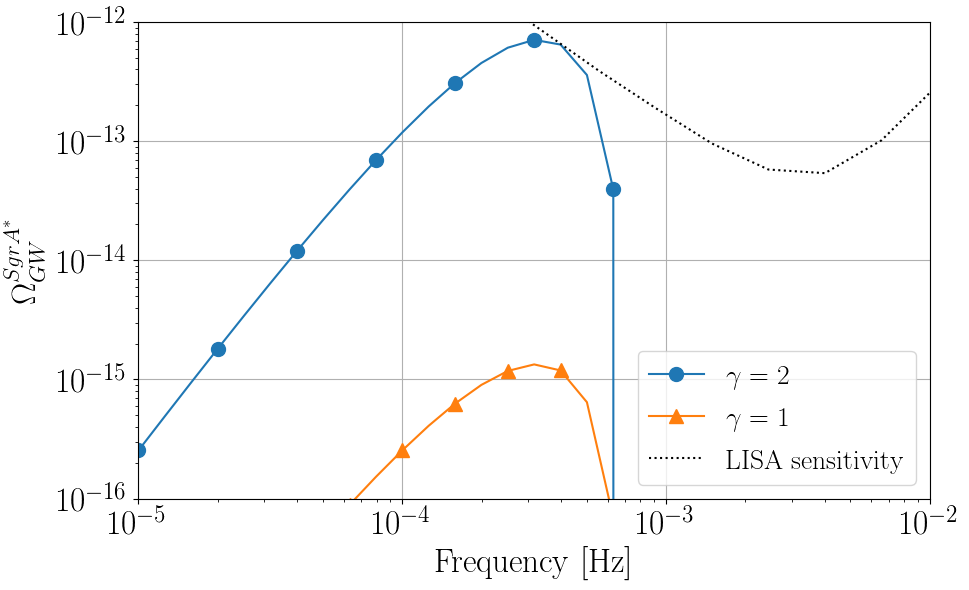} 
   \caption{The \ac{SGWB} energy density spectra of primordial \ac{BH} \acp{EMRI} surrounding Sagittarius A$^\ast$, i.e., the massive \ac{BH} at the center of Milky Way. 
   The mass of primordial \ac{BH} is chosen to be $m_\mathrm{PBH} = 1~ M_\odot$ and the abundance in dark matter is $f_\mathrm{PBH}$ = $10^{-8}$.
   For the initial dark-matter profile power index, $\gamma=2$ and $\gamma=1$ are chosen for illustration.
   Both \ac{SGWB} results peak at the frequency $3\times 10^{-4}$ Hz which is determined by the mass of the massive \ac{BH}.
   The amplitude with $\gamma=2$ is larger than  $\gamma=1$ by three orders of magnitude, inheriting from the three order of magnitude difference of the dark-matter halo density.
   The LISA sensitivity curve for detecting isotropic \ac{SGWB} is also plotted for a qualitative comparison since the \ac{SGWB} from Sagittarius A$^\ast$ is a directional signal.
   The sensitivity curve is for the optimal configuration of LISA which has six links with 5 million km arm length and a 5 year mission duration.}
   \label{fig:SgrA-Omega-GW}
\end{figure}

The mass of  Sagittarius A$^\ast$ located at the center of the Milky Way is $\sim 4\times 10^6 ~M_\odot$ and the distance $d_L$ is $\sim 8$ kpc.
Substitute the values to \cref{Eq:SgrA}, the \ac{SGWB} result is shown in \cref{fig:SgrA-Omega-GW}  with $m_\text{PBH} = 1 M_\odot$ and $f_\text{PBH} = 1\times 10^{-8}$.
Note that the term $m_\text{PBH} f_\text{PBH}$ in \cref{Eq:SgrA} only serves as an overall scaling factor.
We showcase two choices of the  initial dark-matter profile power index, $\gamma=1$ and $2$, respectively.
As a comparison, the LISA sensitivity curve of \ac{SGWB} for the six links, 5 million km arm length configuration and 5 year mission duration (the optimal case) is plotted \cite{LISAscience-phase,LISAscience-inflation}.

The figure shows that the amplitude of \ac{SGWB} with $\gamma=2$ is larger than that with $\gamma=1$ by three orders of magnitude, inheriting from the three order of magnitude difference of the dark-matter halo density as shown in \cref{fig:profile}.
Another feature is that the \ac{SGWB} peaks at $3\times 10^{-4}$ Hz, which is determined by the mass of the central massive \ac{BH} given the dark-matter spike distribution.

It should also be noted that the \ac{SGWB} from Sagittarius A$^\ast$ is a directional signal, but the sensitivity curve is for isotropic \ac{SGWB}.
Therefore the comparison only serves as an order of magnitude estimation due to the lack of a sensitivity curve of space-based detectors for a directional background, but note that the technique of \ac{GW} radiometer and the sensitivity for a directional stochastic background for LIGO have been proposed and calculated by Refs.~\cite{Ballmer:2005uw}.
The search for persistent \ac{GW} signals from pointlike sources has been performed by Advanced \cite{directional1,directional2} and Initial LIGO and Virgo \cite{directional3}. 
Especially, using the data from the first and second observational runs of Advanced LIGO and Virgo, the upper limits for \ac{GW} strain amplitude have been given for three sources with targeting directions: Scorpius X-1, Supernova 1987 A and the Galactic Center, respectively.
Refs.~\cite{space-ansgwb1, space-ansgwb2} have considered the anisotropic \ac{SGWB} search with space-based detectors.
However, a specialized investigation for directional \ac{SGWB} from primordial \ac{BH} \acp{EMRI} at the Galactic center has not been achieved in the context of space-based \ac{GW} detectors.
We thus leave the relevant study as a future work.

\section{The Stochastic Gravitational-Wave Background from Extragalactic Sources} \label{sec:extragalactic}

\subsection{Massive Black Hole Population}

To model the massive \ac{BH} population throughout the cosmic history, we employ the dark-matter halo mass function and the $M_\text{MBH}-M_\text{vir}$ relation which is characterized by \cref{Eq:MBH}.
We choose the halo mass function in Ref.~\cite{Reed:2006rw} which calibrates with a suite of cosmological N-body simulations and takes the finite box size and the cosmic variance effects into account. 
In the actual practice this halo mass function  is generated through invoking the \texttt{Reed07} model in the python package \texttt{hmf} (an acronym for ``halo mass function'') \cite{Murray:2013qza}, where the cosmological parameters are set to be the value of the \textit{Planck} satellite's 2018 results \cite{Planck2018}.
\texttt{Reed07} model is shown to be valid up to redshift $z\sim30$ and for halos with masses $10^{5-12} h^{-1}M_\odot$ \cite{Reed:2006rw}, which is sufficient for our purpose. 

\begin{figure}[htbp] 
   \centering
   \includegraphics[width=\columnwidth]{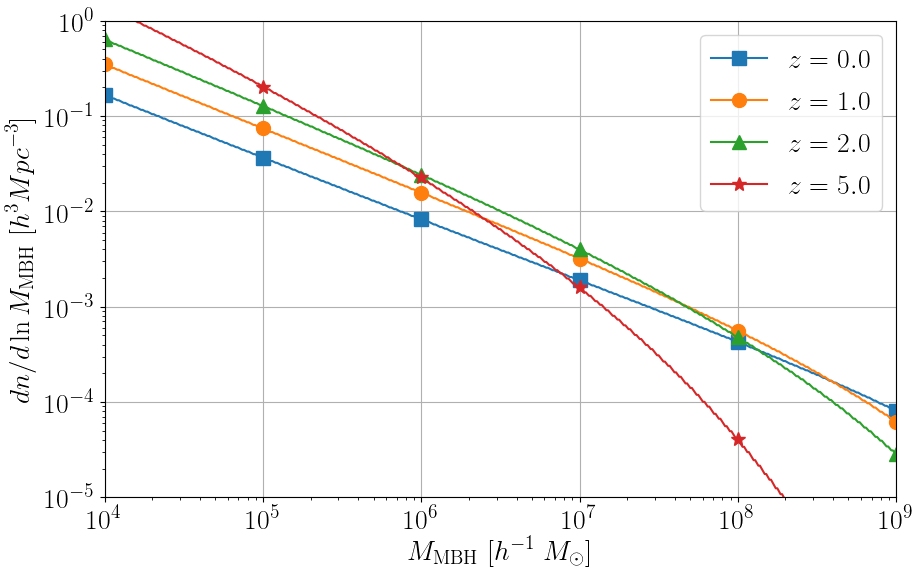} 
   \caption{The solid curves represent the number density of massive \ac{BH} at different redshifts which is derived from \texttt{Reed07} dark-matter halo mass function.
   We will consider the \ac{SGWB} spectrum from the massive \acp{BH} in the mass range $[10^4, \sim10^9]~h^{-1}M_\odot$ as a fiducial model.
   }
   \label{fig:bhmass}
\end{figure}

\cref{fig:bhmass} shows our results of the number density $dn/d\ln M_\textrm{MBH}$ for massive \acp{BH} in different redshifts. 
Since astronomical observations have confirmed the existence of massive \acp{BH} with $\sim10^{4}~M_\odot$ (for example see Ref.~\cite{SMBH1e-4}), we will consider the \ac{SGWB} spectrum from the massive \acp{BH} in the mass range $[10^4, \sim10^9]~h^{-1}M_\odot$ as a fiducial model.
The upper mass limit is determined by the condition that the emitted \acp{GW} are in the LISA sensitive band and $\sim10^9~ M_\odot$ is sufficient for this purpose.

\subsection{Stochastic Gravitational-Wave Background Spectrum}

The complete \ac{SGWB} contribution can be obtained by taking the sum from the \acp{EMRI} of all extragalactic massive \acp{BH}
\be\label{Eq:finalOmega}
\Omega_\textrm{GW}(\nu) = \frac{\nu}{\rho_c}\frac{4\pi}{c}\iint\frac{dF_\text{GW}}{d\nu}\frac{dn}{dM_\text{MBH}}dM_\text{MBH}\chi^2 d\chi,
\ee
where $dn/dM_\text{MBH} $ is the number density of massive \acp{BH} and $\chi$ is the sources' comoving distance. Combining \cref{Eq:dFoverdnu} and \cref{Eq:finalOmega} yields
\begin{align}\label{Eq:omegacontinuous}
\Omega_\text{GW}(\nu) &= f_\text{PBH} m _\text{PBH} \frac{\nu}{\rho_c} \frac{256 \pi^2 G^{7/2} }{15 c^7}\\  \nonumber
&\times \int \frac{dz}{(1+z)H(z)} 
\int \frac{\rho_\text{DM}M_\text{MBH}^{5/2}}{r^{1/2}}   \frac{dn}{dM_\text{MBH}} dM_\text{MBH}.
\end{align}
As a fiducial model, we consider the sources in the redshift range $[0,5]$.
Contributions from higher redshift can be negligible as will be indicated in \cref{sec:robust}.

\begin{figure}[htbp]
   \centering
   \includegraphics[width=\columnwidth]{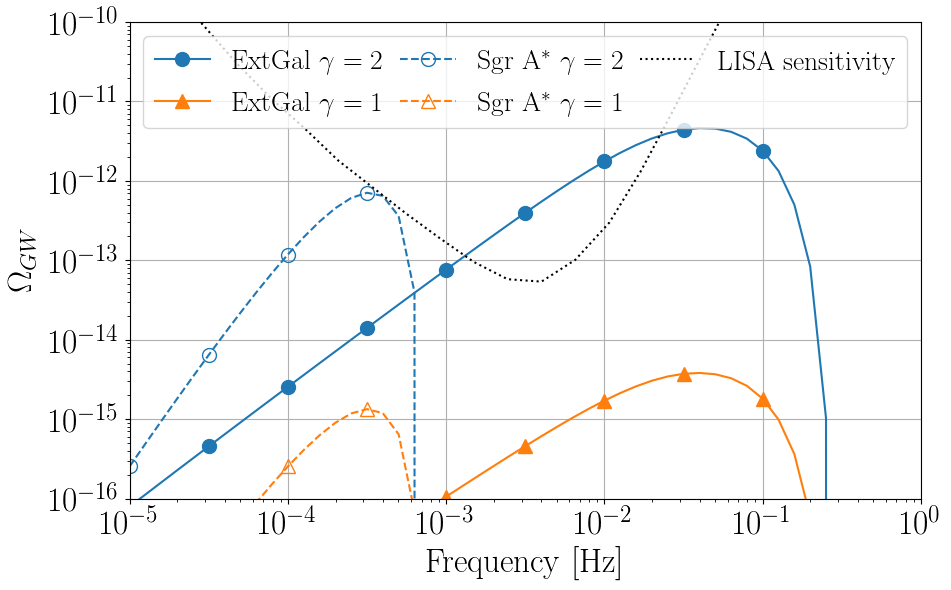}
\caption{The \ac{SGWB} spectra from primordial \ac{BH} \acp{EMRI} surrounding Sagittarius A$^\ast$ (``Sgr A$^\ast$'') and the extragalactic massive \acp{BH} (``ExtGal'') in the redshift range $[0,5]$.
The mass of the primordial \ac{BH} is assumed to be $1 M_\odot$ and the abundance in dark matter is $10^{-8}$. 
The dark-matter initial profile power index is chosen to be $\gamma=1$ and $\gamma=2$, respectively.
The $\gamma=2$ extragalactic \ac{SGWB} already reaches the detectable zone of LISA.
 }\label{fig:ExtGal}
\end{figure}

With  $m_\text{PBH} = 1 M_\odot$ and $f_\text{PBH}=10^{-8}$, the results of extragalactic \ac{SGWB}  with $\gamma=2$ and $\gamma=1$  are shown in \cref{fig:ExtGal}.
The results from Sagittarius A$^\ast$ are also replotted for comparison.
The extragalactic \ac{SGWB} peaks at a higher frequency ($\sim4\times 10^{-2}$ Hz), due to the contributions from \acp{BH} less massive than Sagittarius A$^\ast$.

Note that the extragalactic \ac{SGWB} is an isotropic signal, and thus can be compared with the \ac{LISA} sensitivity curve directly. It shows that the amplitude of extragalactic \ac{SGWB}  for $\gamma=2$ already reaches the detectable zone, which means that the future LISA searching results can probe the abundance of $1 M_\odot$ primordial \acp{BH} to be as low as $\sim 10^{-9}$ for the $\gamma=2$ case.
The amplitude of $\gamma=1$ \ac{SGWB} is smaller by three orders of magnitude due to a smaller dark-matter density profile.

The main uncertainty comes from the dark-matter initial profile power index. 
Future astrophysical progress on understanding the dark-matter profile at galactic center can shed light on a more robust prediction on the primordial \ac{BH} \acp{EMRI} event rate. 
Conversely, future \ac{GW} search with space-based detectors can also be beneficial for study for the dark-matter profile.

\subsection{Density Enhancement due to Gravitational-Wave Dissipation}

Another consideration is that the orbit of inspiraling primordial \acp{BH} will gradually decay due to \ac{GW} dissipation. 
As a consequence, the primordial \ac{BH} density profile gets further concentrated.
To quantify this effect, we notice that astrophysical spectroscopy surveys have confirmed that the first galaxies form at redshift $z \gtrsim10$ ($\sim$13 Gyr from now) \cite{FirstGalaxy}.
Therefore we make the extreme assumption that all the primordial \acp{BH} \acp{EMRI} start to decay at $z =10$, thus the decay duration equals to  $\sim$13 Gyr, to obtain an upper limit of the dark-matter density profile after concentration.

The relation among the orbital decay duration $\Delta t$, the final orbit radius $r_f$ and the initial radius $r_i$ is given by Refs.~\cite{peter2,peter1} as follows
\be\label{eq:radiusdecay}
\Delta t = \frac{5}{256} \frac{c^5}{G^3}\frac{1}{\eta M_{tot}^3}(r_i^4-r_f^4). 
\ee

Let $\Delta t = 13$ Gyr and employ the mass conservation relation \footnote{We have numerically verified that the mass loss due to \ac{GW} dissipation is negligible by comparing the \ac{GW} energy with the mass energy of primordial \acp{BH}.}
\be 
\rho_f(r_f) = \frac{r_i^2}{r_f^2} \rho_i(r_i),
\ee
the enhanced density profile $\rho_f$ can be determined from the initial dark-matter profile $\rho_i$ which is assumed to the spike distribution.

For comparison, we also consider a less optimistic case where the orbital decay duration $\Delta t$ is an average of a flat distribution which  ranges from zero to 13 Gyr, i.e., the primordial \ac{BH} final orbital radius after \ac{GW} dissipation $r_f$ can be determined  by $r_i$ with the following expression
\be
r_f =\left.{\int_0^{13~\mathrm{Gyr}} \left(r_i^4-\frac{256}{5}\frac{G^3}{c^5}\eta M_{tot}^3{\Delta t}\right)^{1/4} \text{d} \Delta t} \middle/ \right.13~\mathrm{Gyr}.
\ee
However, we find numerically that the relative difference of the resulting \ac{SGWB} spectra between the most optimistic case (i.e., $\Delta t$=13 Gyr) and the less-optimistic case is negligibly small and at most $\mathcal{O}(10\%)$; therefore we will only consider the most optimistic case in the following as a representative of the \ac{GW} dissipation effect.

\begin{figure}[htbp]
   \centering
   \includegraphics[width=\columnwidth]{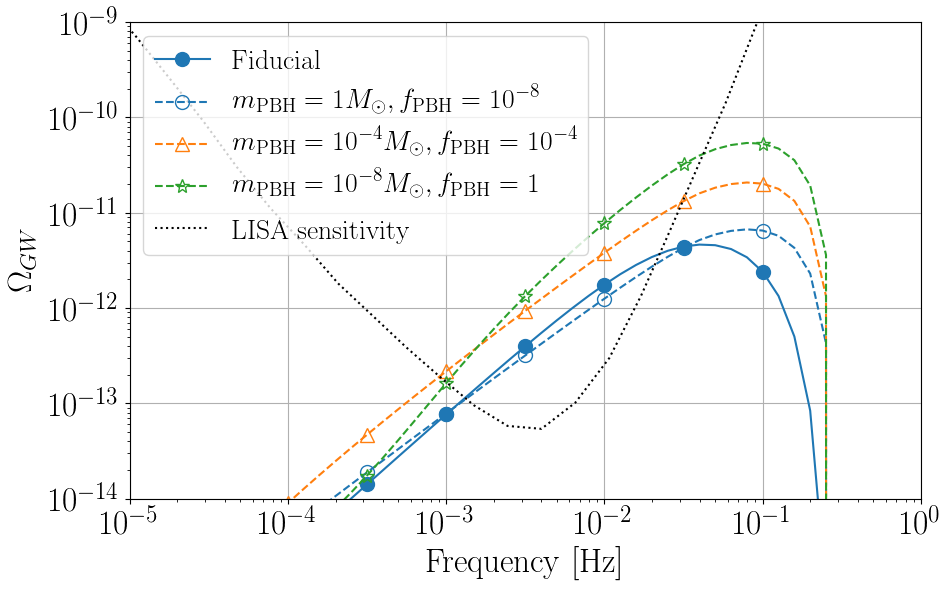}
\caption{ A comparison between the \ac{SGWB} spectra with and without primordial \ac{BH} density enhancement effect due to \ac{GW} dissipation.
The fiducial model is calculated from the $\gamma=2$ dark-matter spike distribution for $m_\mathrm{PBH} = 1 M_\odot, f_\mathrm{PBH} = 10^{-8}$.
The three dashed lines are from the $\gamma=2$ enhanced dark-matter spike distribution due to \ac{GW} dissipation.
The mass and abundance parameters are so chosen to keep $m_\mathrm{PBH}f_\mathrm{PBH} = 10^{-8}$.
}\label{fig:GWenhance}
\end{figure}

Choosing $\gamma=2$, the fiducial \ac{SGWB} result without density enhancement for $m_\mathrm{PBH} = 1 M_\odot, f_\mathrm{PBH} = 10^{-8}$ and the \ac{SGWB} spectra with density enhancement for $m_\mathrm{PBH}=1 M_\odot,10^{-4} M_\odot$ and $10^{-8} M_\odot$, respectively, are shown in \cref{fig:GWenhance}.
For the enhanced \ac{SGWB} results, the primordial \ac{BH} abundance is modified accordingly to keep $m_\mathrm{PBH}f_\mathrm{PBH} = 10^{-8}$.
Since \ac{GW} dissipation makes primordial \acp{BH} get concentrated and closer to the center, the amplitude of \ac{SGWB} spectra gets boosted and the frequency of the peak changes to $\sim 10^{-1}$ Hz.
As the mass of primordial \acp{BH} decreases, the amplitude boost becomes more significant.
The amplitude for $m_\mathrm{PBH} = 10^{-8} M_\odot$ is larger than the fiducial result by one order of magnitude.
This can be expected from \cref{eq:radiusdecay} that a smaller value of symmetric mass ratio $\eta$ leads to a more significant orbital decay.

\subsection{Modeling Systematics of the Massive Black Hole Population}\label{sec:robust}

To investigate the systematics on modeling the extragalactic massive \ac{BH} population, we vary the redshift integral limits, the lower mass of massive \acp{BH} and the massive \ac{BH} population model, respectively, and plot the corresponding \ac{SGWB} spectra.
The results are shown in \cref{fig:sgwbspace-robustness}. 
All \ac{SGWB} spectra are calculated from $\gamma=2$ dark-matter spike distribution and $f_\text{PBH} = 10^{-8}$, $m_\text{PBH} = 1 M_\odot$.

\subsubsection{The Redshift Limits of the Integral}

The left column of \cref{fig:sgwbspace-robustness} shows the \ac{SGWB} component contributions from different redshift ranges, $[0,1], [1,3]$, and $[3,5]$, respectively.
The fiducial result is the sum, i.e., obtained from the redshift range $[0,5]$.
We can see that the contribution from $z\in [3,5]$ is subdominant and accounts for at most $~10\%$ of the fiducial result.
We thus conclude that the choice of $5$ for the redshift upper limit is sufficient to capture the dominant contribution.

\subsubsection{The Lower Mass Cutoff of the Massive Black Holes}

As mentioned, we chose $10^4 ~h^{-1} M_\odot$ to be the minimum mass of massive \acp{BH} to account for the existence of such massive \acp{BH} from astronomical observations.
Nevertheless we change this value to $10^5 ~h^{-1} M_\odot$ to investigate the contribution from different mass ranges.
The result is shown in the middle column of \cref{fig:sgwbspace-robustness}.
Compared with the fiducial result whose lower mass cutoff is $10^4~h^{-1} M_\odot$, the result shows that the \ac{SGWB} contributed by $[10^4,10^5]~h^{-1} M_\odot$ massive \acp{BH} is mainly in the relative high frequency range $[\sim10^{-2}, \sim10^{-1}]$ Hz.
Therefore the shape of the \ac{SGWB} spectrum, especially the frequency of the peak, can provide information of the underlying massive \ac{BH} population. 

\subsubsection{The Population Model of Massive Black Holes}

\begin{figure*}[htbp]
   \centering
   \includegraphics[width=2\columnwidth]{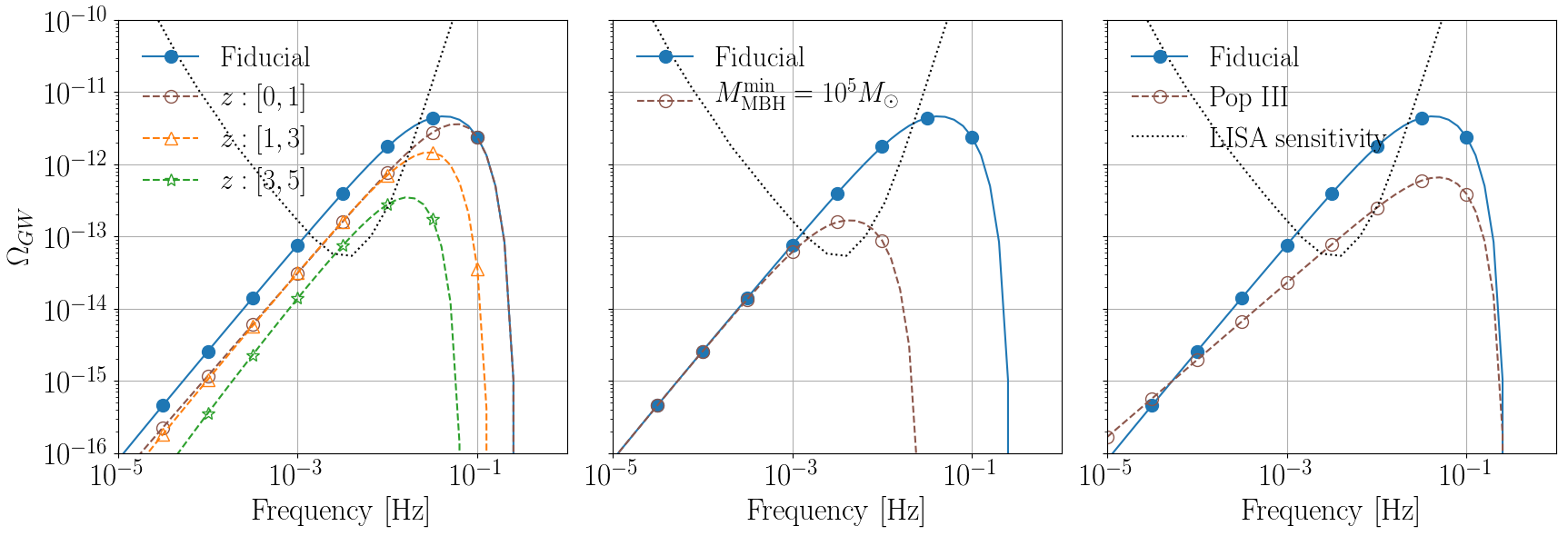}
\caption{The effect on \ac{SGWB} spectra from the modeling systematics of massive \ac{BH}. 
The three figures present the \ac{SGWB} by varying the redshift integral limits, the lower mass of massive \acp{BH} and the massive \ac{BH} population model, respectively.
All \ac{SGWB} spectra are calculated from $\gamma=2$ dark-matter spike distribution and $f_\text{PBH} = 10^{-8}$, $m_\text{PBH} = 1 M_\odot$.}
\label{fig:sgwbspace-robustness}
\end{figure*}

A third investigation is to substitute the massive \ac{BH} population model derived from the \texttt{Reed07} dark-matter halo mass function to that proposed by Refs~.\cite{Barausse:2012fy,Sesana:2014bea,Antonini:2015cqa,Antonini:2015sza} accounting for the massive \acp{BH} formed from population III star seeds.
The number density of massive \acp{BH} of this model is
\be\label{eq:mbhpopiii}
\frac{dn}{d\log M_\mathrm{MBH}} = 0.005\( \frac{M_\mathrm{MBH}}{3\times 10^6 M_\odot}\)^{-0.3} \mathrm{Mpc^{-3}}
\ee 
The right column of \cref{fig:sgwbspace-robustness} presents the results.
It shows that the amplitude of the peak is one order of magnitude smaller than the fiducial model.
This is because the number density of massive \acp{BH} with mass $[10^4,10^6]~h^{-1} M_\odot$ from \cref{eq:mbhpopiii} is less than that from the \texttt{Reed07} derived population.

The above investigations give a  quantitative measurement about the effect on \ac{SGWB} spectra from different modeling systematics of extragalactic massive \ac{BH}, which can be improved in the future once a better understanding of the population of massive \acp{BH} is obtained.
In addition, the future \ac{SGWB} search with space-based detectors can serve as a tool to probe the population information of massive \acp{BH}.

\section{Constraints on Primordial Black Hole Abundance}\label{sec:PBH-EMRI-constraint}

\begin{figure}[htbp]
   \centering
   \includegraphics[width=\columnwidth]{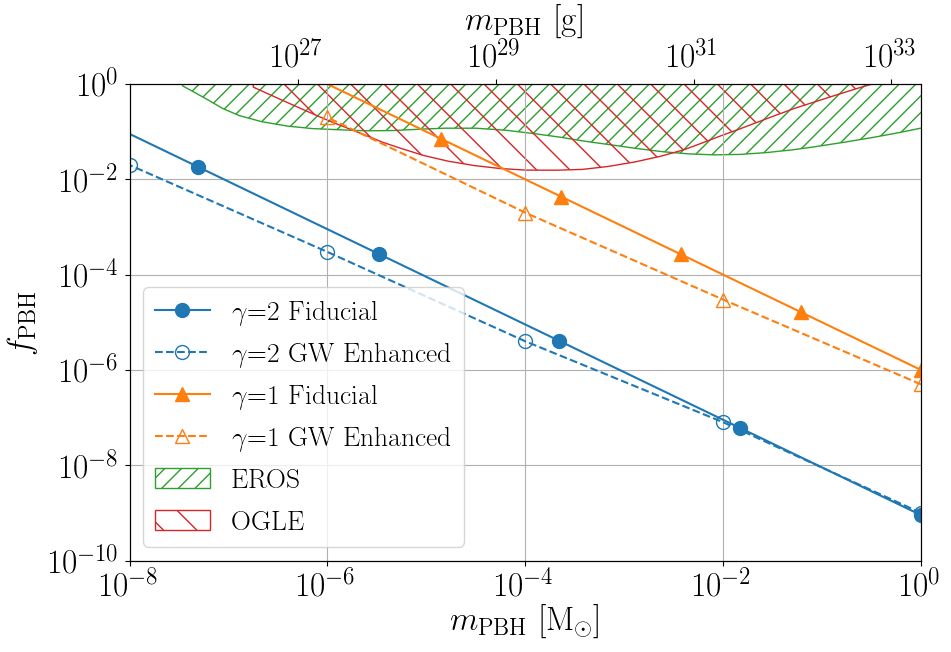}
   \caption{The projected constraints on primordial \ac{BH} abundance in dark matter with \ac{LISA} for $\gamma=1$ and $\gamma=2$, with and without the enhancement due to \ac{GW} dissipation, respectively.
   As a comparison, we also plot the current constraint from the microlensing search collaborations OGLE \cite{OGLE2019} and EROS \cite{EROS2006}.
   }\label{Fig:constraints}
\end{figure}

By comparing the \ac{SGWB} spectrum with the LISA sensitivity and applying the condition
\be
\Omega_\text{GW}(\nu; m_\text{PBH}, f_\text{PBH} ) \leq \Omega_\text{GW}^\text{LISA sensitivity},
\ee
the upper limit on primordial \ac{BH} abundance $f^\text{max}_\text{PBH}$ can be obtained for a null searching result.
In \cref{Fig:constraints}, we plot the results for different primordial \ac{BH} masses from the models of $\gamma=1$, $\gamma=2$, with and without the enhancement due to \ac{GW} dissipation, respectively.
As a comparison, we also plot the current constraint from the microlensing search collaborations OGLE \cite{OGLE2019} and EROS \cite{EROS2006}.

It shows that LISA can probe the primordial \ac{BH} abundance in a large range of masses, $[10^{-9}, 1] M_\odot$ for  $\gamma=2$ and $[10^{-6}, 1] M_\odot$ for  $\gamma=1$, respectively. 
We do not consider more massive primordial \acp{BH} because it may break the condition of extreme mass ratio.
For primordial \acp{BH} with $1 M_\odot$, LISA can constrain the abundance to be $\sim10^{-6}$ ($\gamma=1$) or $\sim10^{-9}$ ($\gamma=2$). 
The enhancement effect due to \ac{GW} dissipation can further improve the constraint at the lower end of the mass range.
This would be the most sensitive method proposed and quantified by now for detecting or constraining subsolar mass primordial \acp{BH}.

\section{Conclusions}\label{sec:PBH-EMRI-conclusion}

In this work we have investigated the \ac{SGWB} from primordial \ac{BH} \acp{EMRI} surrounding Sagittarius A$^\ast$ and the extragalactic massive \acp{BH}, respectively, expanding the previous work of Ref.~\cite{Kuhnel:2017bvu, SgrA} by including the complete extragalactic contribution.
After modeling the event rate, the \ac{SGWB} energy density spectra are calculated.
The contributions from Sagittarius A$^\ast$ and extragalactic massive \acp{BH} peak at different characteristic frequencies.
Future space-based \ac{GW} detectors such as LISA may utilize this feature to give decisive evidence about the existence of primordial \ac{BH} dark matter at the galactic center.
Finally, LISA can also constrain the abundance of primordial \ac{BH} in dark matter if there will be a null \ac{SGWB} searching result.
For a NFW induced dark-matter spike profile ($\gamma=1$), LISA can exclude the existence of $1 M_\odot$ primordial \ac{BH} with any abundance greater than $10^{-6}$ of dark matter.
A steeper dark-matter profile power index $\gamma=2$ can make the constraint even tighter by three orders of magnitude.
This renders \acp{GW} in the space-based frequency band  a powerful tool to hunting for subsolar mass primordial \acp{BH}.
However, modeling uncertainties exist from the dark-matter spike profile power index and the extragalactic massive \acp{BH} population as quantified in \cref{sec:extragalactic}.
Future astrophysical progress on understanding these modeling systematics can help further improve the ability of the \ac{GW} window to search for primordial \acp{BH}.
In addition, as a future work, we will apply the specialized technique of the \ac{GW} radiometer on the LISA detector to investigate the detection ability for the directional \ac{EMRI} background from Sagittarius A$^\ast$.
It would also be interesting to study whether the stochastic background is continuous or popcorn \cite{Rosado:2011kv, Regimbau:2009av}, thereby applying different optimal detection strategies accordingly \cite{PhysRevX.8.021019}.

\begin{acknowledgements}
Q.-G.H. is supported by grants from NSFC (Grants No. 11690021, No. 11975019, No. 11947302, No. 11991053), the Strategic Priority Research Program of Chinese Academy of Sciences (Grants No. XDB23000000, No. XDA15020701), and Key Research Program of Frontier Sciences, CAS, Grant No. ZDBS-LY-7009. T.G.F.L. is partially supported by grants from the Research Grants Council of the Hong Kong (Project No. 24304317), Research Committee of the Chinese University of Hong Kong and the Croucher Foundation of Hong Kong.
\end{acknowledgements}

\bibliography{emri-pbh.bib}

\end{document}